\documentclass[prb,preprint,superscriptaddress,onecolumn]{revtex4-1} 

\usepackage{amsmath}  
\usepackage{amsfonts} 
\usepackage{graphicx} 
\usepackage{esvect}
\usepackage{mathtools}
\usepackage{subcaption}
\usepackage{xcolor}
\usepackage{appendix}

\usepackage{siunitx} 
\usepackage{placeins}

\usepackage{float}

\usepackage{pbox}

\usepackage{url}

\usepackage{xcolor}
\pagecolor{white}

\def\Z{\mathbb{Z}}

\begin{document}


\title{Equal Radiation Frequencies from Different Transitions
in the Non-Relativistic Quantum Mechanical Hydrogen Atom}

\author{Tuan K. Do}
\email{kdo@princeton.edu}
\affiliation{Department of Mathematics, Princeton University, Princeton, NJ 08544, USA}

\author{Trung V. Phan}
\email{trung.phan@yale.edu}
\affiliation{Department of Molecular, Cellular, and Developmental Biology, Yale University, New Haven, CT 06520, USA}

%
%
%
%
%

\begin{abstract}
Is it possible that two different transitions in the non-relativistic quantum mechanical model of the hydrogen atom give the same frequency? That is, can different energy level transitions in a hydrogen atom have the same photon radiation frequency? This question, which was asked during a Ph.D. oral exam in 1997 at the University of Colorado Boulder, is well-known among physics graduate students. We show a general solution to this question, in which all equifrequency transition pairs can be obtained from the set of solutions of a Diophantine equation. This fun puzzle is a simple yet concrete example of how number theory can be relevant to quantum systems, a curious theme that emerges in theoretical physics but is usually inaccessible to the general audience.
\end{abstract}

\maketitle

\date{\today}

\section{Introduction}

For more than a century, quantum mechanics has been the fundamental theory that guides our understanding of how nature works at the scale of atoms and subatomic particles. The hydrogen atom is the simplest possible atom for theoretical investigation, consisting of only a single proton and a single electron orbiting around \cite{lakhtakia1996models}. While the hydrogen atom has been studied intensively since the dawn of quantum mechanics as a demonstration of what measurements can reveal about atoms, there are still surprises and hidden structures \cite{mills2000hydrogen}. One of such is the emergence of equifrequency transitions, in which many distinctive jumps between atomic levels can radiate identical photon energy. This question was raised during a Ph.D. oral exam in 1997 at the University of Colorado Boulder \cite{question} and soon became well-known in the physics community, especially among graduate students. The answer is definitely yes, and infinitely many transitions have been found \cite{answer}, but to the best of our knowledge, a generalization is still lacking.
\begin{figure}[!htb]
\centering
\includegraphics[width=0.7\textwidth]{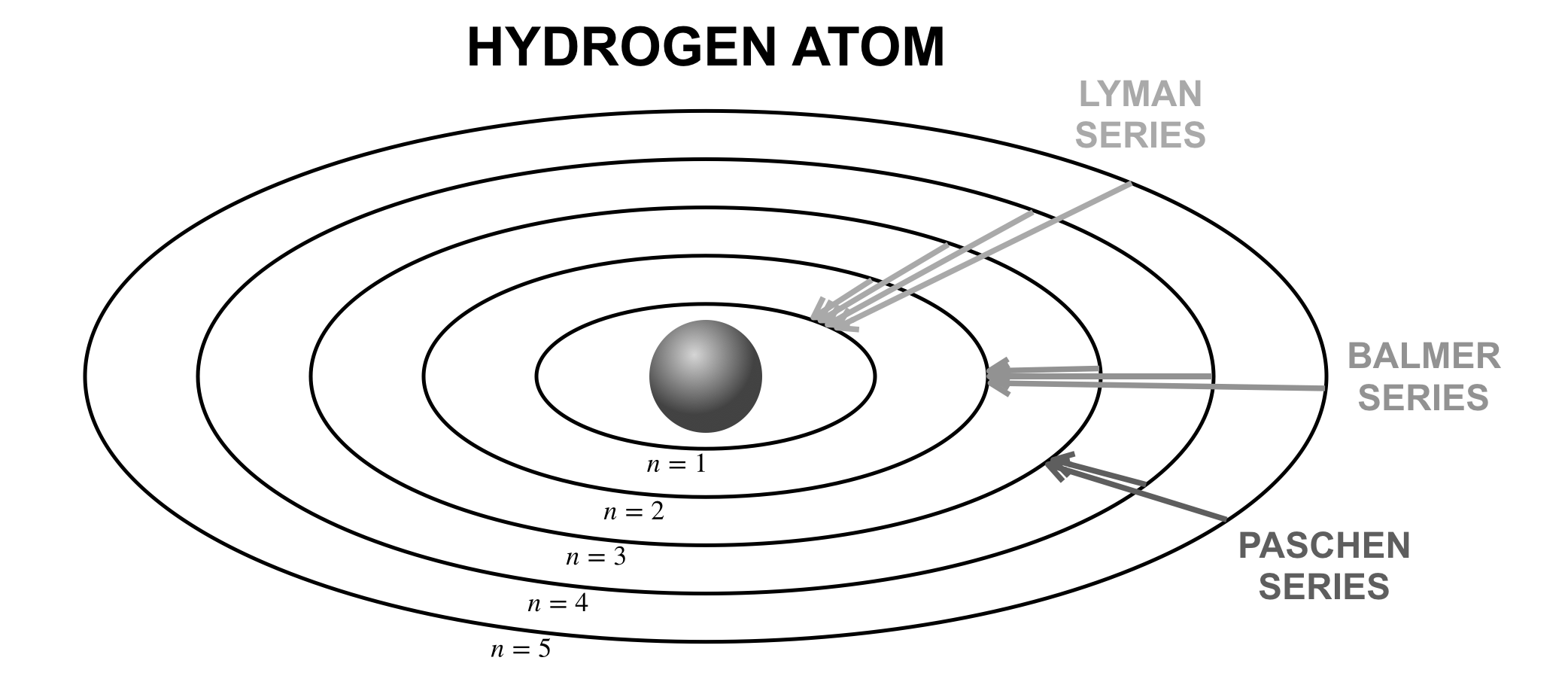}
\caption{Energy level transitions in the non-relativisitic quantum mechanical model of the hydrogen atom.
An electron jumps from an outer ring $n_1$-th to an inner ring $n_2$-th, emits a photon with radiation energy $\Delta E \propto n_2^{-2} - n_1^{-2}$.}
\label{fig1}
\end{figure}

Here we show the connection between the above question with a Diophantine equation \cite{Diophantine} and present a general solution, {\it i.e.} how all equifrequency transition pairs can be obtained. While this finding might not address any foundational issue or important problem in quantum mechanics, it definitely provides us with a more complete understanding of the most popular atom in all quantum mechanics textbooks, on the relation between atomic levels (disregarding degeneracies due to angular momentum and spin). It is also a simple illustration of how number theory can be of relevance to physics \cite{book}, in a way accessible to non-experts.

\section{A General Solution for All Equifrequency Transitions}

In quantum mechanics, the $n$-th energy level of a hydrogen atom is given by $E(n) = - E_o/n^2$, where $n \in \mathbb{Z}^+$ is a positive integer and $E_o = 13.6$ eV is the Rydberg energy \cite{Griffiths_Intro}. For simplicity, we will not consider any relativistic effects \cite{auvil1978relativistic} or other corrections (such as fine structure \cite{lamb1950fine,lamb1951fine}) to this equation. The challenge is to find all transition pairs ($n_1 \rightarrow n_2$,$n_3 \rightarrow n_4$) with equal radiation energy, which means,
\begin{equation}
\centering
\boxed{\frac1{n_2^2} - \frac1{n_1^2} = \frac1{n_4^2} - \frac1{n_3^2} > 0 }\ \ . \ \
\label{main}
\end{equation}
Here we will find a general solution of this equation, including trivial solutions where $n_1=n_3$ and $n_2=n_4$.

Consider the Diophantine equation \cite{Diophantine} with a parameter $s \in \mathbb{Z}^+$ and unknowns $x,y,z \in \mathbb{Z}^+$,
\begin{equation}
x^2 - y^2 = sz^2 \ \ . \ \
\label{pell_like}
\end{equation} 
With any two solutions $(x_1,y_1,z_1)$ and $(x_2, y_2, z_2)$ to this equation, for any positive integer pair $(t_1,t_2)$ that satisfies
\begin{equation}
x_1 y_1 t_1 z_2 = x_2 y_2 t_2 z_1 \ ,
\label{product}
\end{equation} 
a solution to equation (\ref{main}) can be obtained:
\begin{equation}
(n_1,n_2,n_3,n_4) = (x_1 t_1, y_1 t_1, x_2 t_2, y_2 t_2) \ \ , \ \ 
\label{solution}
\end{equation} 
which can be checked by direct substitution. 
To generate all solutions $(t_1,t_2)$ to equation (\ref{product}), we use any $k \in \mathbb{Z}^+$ and $G=\text{gcd}(x_1 y_1 z_2,x_2 y_2 z_1)$,
\begin{equation}
t_1 = kx_2 y_2 z_1/G \ \ , \ \ t_2 = kx_1 y_1 z_2/G \ \ , \ \
\label{get_tt}
\end{equation} 
where the operation $\text{gcd}(\alpha , \beta)$ determines the greatest common divisor of $\alpha,\beta \in \mathbb{Z}^+$ \cite{long1987elementary}.

We can prove that the above procedure comprises all solutions of equation (\ref{main}). 
Start from this equation, denote $t_1' = \text{gcd}(n_1,n_2)$ and $t_2' = \text{gcd}(n_3,n_4)$. Write $n_1=x_1't_1',n_2=y_1't_1',n_3=x_2't_2',n_4=y_2't_2'$. 
Note that $n_1>n_2$ and $n_3>n_4$ i.e. $x_1'>y_1'$ and $x_2'>y_2'$. 
Then, we rewrite (\ref{main}) as,
\begin{equation}
    \frac{x_1'^2-y_1'^2}{x_2'^2-y_2'^2}=\left(\frac{x_1'y_1't_1'}{x_2'y_2't_2'}\right)^2
\end{equation}
and put the fraction $x_1'y_1't_1'/x_2'y_2't_2'$ into irreducible form $z_1'/z_2'$ where $\text{gcd}(z_1',z_2') = 1$ and both are nonzero,
\begin{equation}
    \frac{x_1'y_1't_1'}{x_2'y_2't_2'} = \frac{z_1'}{z_2'} \ \ . \ \
\label{fraction}
\end{equation}
Thus,
\begin{equation}
    \frac{x_1'^2-y_1'^2}{x_2'^2-y_2'^2}=\frac{z_1'^2}{z_2'^2}
\end{equation}
and hence there exists $s' \in \Z^+$ such that,
\begin{equation}
    x_1'^2-y_1'^2 =s'z_1'^2 \ \ , \ \ x_2'^2-y_2'^2 =s'z_2'^2.
\label{2ndcondition}
\end{equation}
Notice here that condition (\ref{fraction}) is exactly equation (\ref{product}) and condition (\ref{2ndcondition}) provides us two solutions of (\ref{pell_like}). 
Combine with the above paragraph, we see these two conditions (\ref{fraction}) and (\ref{2ndcondition}) are both necessary and sufficient. This completes the proof.

To generate the set of all nonzero integer solutions $(x,y,z)$ to equation (\ref{pell_like}), we will need the set of all non-zero rational solutions $(a,b)$ to its dehomogenized version (by dividing both sides of (\ref{pell_like}) by $1/y^2$),
\begin{equation}
a^2 - 1 = sb^2 \ \ . \ \
\label{pell_like_rational}
\end{equation} 
This equation is very similar to the Pell equation \cite{Pellsolution}, but can be solved by much simpler method. By taking any $(a,b)=(a_1/a_2,b_1/b_2)$ satisfies (\ref{pell_like_rational}) and any $l \in \mathbb{Z}$, we obtain all triples,
\begin{equation} 
(x,y,z)=\left( \frac{ l a_1 b_2}{G_2}, \frac{l a_2 b_2}{G_2}, \frac{ l a_2 b_1}{G_2} \right) \ \ , \ \ 
\label{get_xyz}
\end{equation}
of (\ref{pell_like}) where $G_2 = \text{gcd}(a_2,b_2)$.

\begin{figure*}[htb]
\centering
\includegraphics[width=0.9\textwidth]{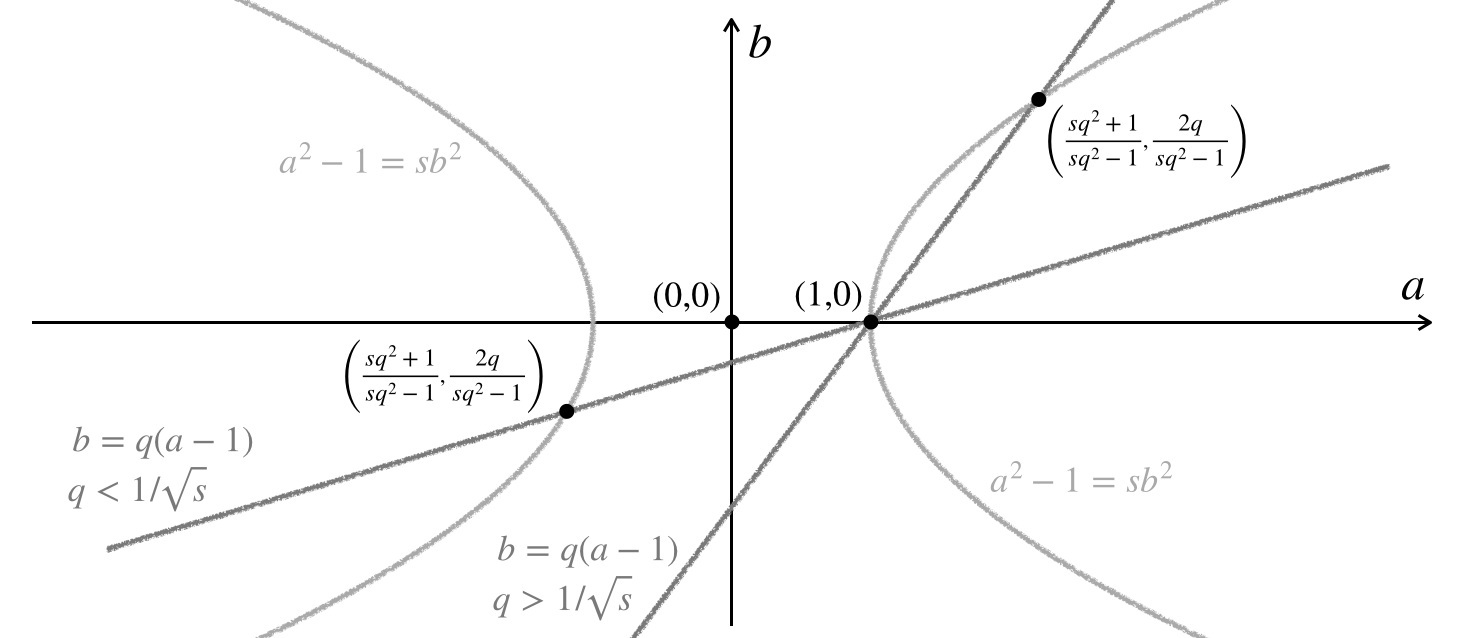}
\caption{The geometric representation of curve equation \eqref{pell_like_rational} and line equation $b=q(a-1)$ in the $a$-$b$ plane. The intersection in the first quadrant gives a solution to equation \eqref{pell_like_rational}.}
\label{fig2}
\end{figure*}

The geometric way \cite{hindry2013diophantine} to deal with equation (\ref{pell_like_rational}) is to draw in the $ab$-plane a line passing through $(1,0)$ with a rational slope $q \in \mathbb{Q}$, say the line $b=q(a-1)$ (see Fig \ref{fig2}). 
For $q^2 \neq 1/s$, this line will cut the curve (\ref{pell_like_rational}) at another point,
\begin{equation}
(a,b) = \left( \frac{sq^2+1}{sq^2-1} , \frac{2q}{sq^2-1} \right) \ \ , \ \ 
\label{ab_solution}
\end{equation}
and more importantly, all solutions of equation (\ref{pell_like_rational}) can be attained this way by varying $q$. 
Note that $q=0$ gives $z=0 \notin \mathbb{Z}^+$ and changing the sign of $q$ changes the sign of $(a,b)$. 
Hence, if we let $q=q_1/q_2$ where $q_1 \in \mathbb{Z}\backslash \{0\}$, $q_2 \in \mathbb{Z}^+$ then $a=a_1/a_2$, $b=b_1/b_2$ where,
\begin{align}
a_1= sq_1^2 + q_2^2 \ \ , \ \ a_2 = sq_1^2 - q_2^2 \ \ , \ \ 
\label{get_a_irr}
\\
b_1 = 2q_1 q_2 \ \ , \ \ b_2 = sq_1^2 - q_2^2 \ \ . \ \ 
\label{get_b_irr}
\end{align}
The positive triple $(x,y,z)$ can be obtained now from (\ref{get_xyz}) with the correct sign choice.

\begin{figure*}[htb]
\centering
\includegraphics[width=0.9\textwidth]{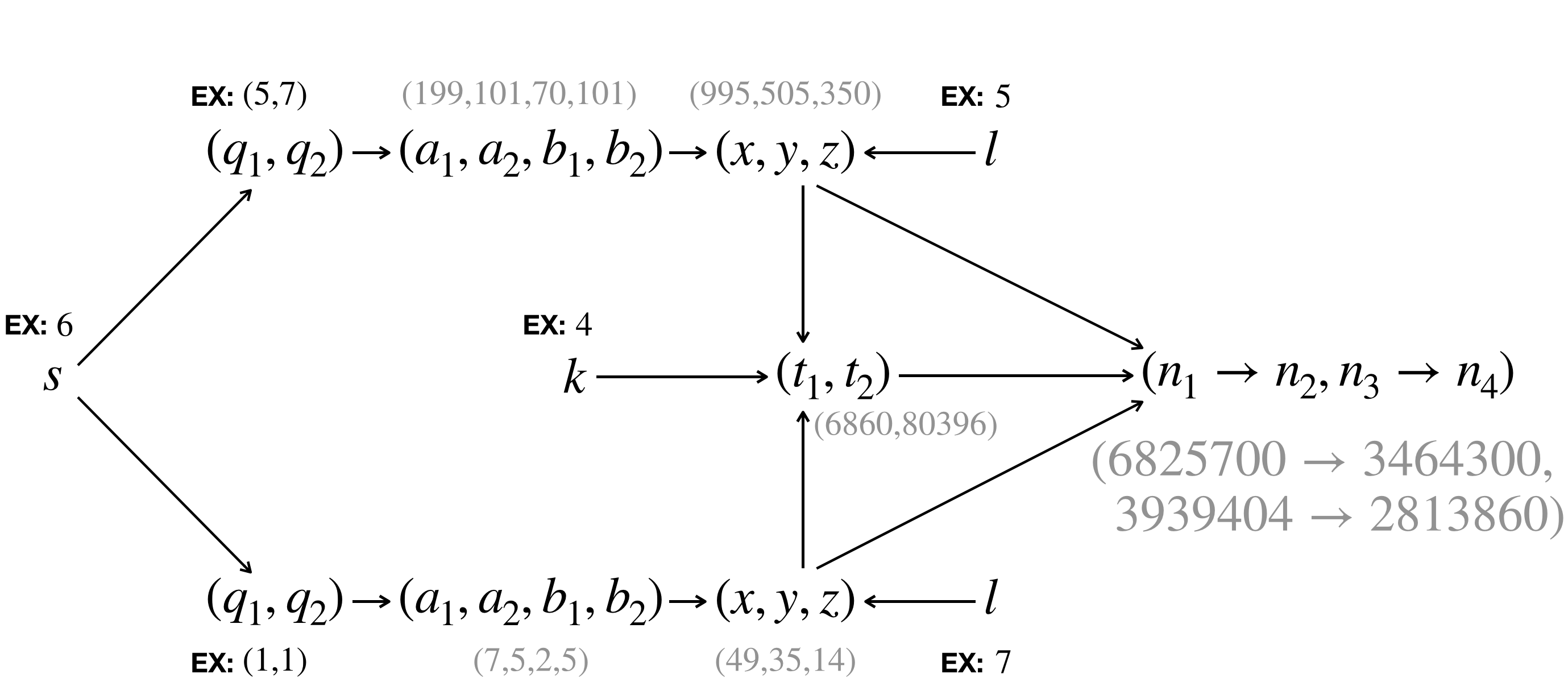}
\caption{A demonstration for the procedure to get an equifrequency transition pair. 
Here we start by selecting $s=6$, then from $(q_1,q_2)=(5,7)$ and $l=5$ we get $(x_1,y_1,z_1)=(995,505,350)$, from $(q_1,q_2)=(1,1)$ and $l=7$ we get $(x_2,y_2,z_2)=(49,35,14)$. 
Then, with $k=4$ we arrive at $(n_1 \rightarrow n_2,n_3 \rightarrow n_4)=(6825700 \rightarrow 3464300,$ $3939404 \rightarrow 2813860)$, which can be checked to satisfy equation (\ref{main}).} 
\label{fig3}
\end{figure*}

In summary, we can generate a solution $(x,y,z)$ to equation (\ref{pell_like}) with parameter $s\in \mathbb{Z}^+$ from any number $q=q_1/q_2\neq 0$. Given the pair, we go through equations (\ref{get_a_irr})-(\ref{get_b_irr}), pick a value $l \in \mathbb{Z}$ and use equation (\ref{get_xyz}) to arrive at $(x,y,z)$. Then with two such solutions, say $(x_1,y_1,z_1)$ and $(x_2,y_2,z_2)$, we pick a value $k \in \mathbb{Z}^+$ and use equation (\ref{get_tt}) to get $(t_1,t_2)$ before plugging in equation (\ref{solution}) to get a pair $(n_1 \rightarrow n_2, n_3 \rightarrow n_4)$. 
See Fig \ref{fig3} for a demonstration. The key difference in our approach compared to previous ones is using (\ref{pell_like}), where we can generate all possible rational solutions, which enables us to find all possible solutions of the puzzle (\ref{main}).

\section{Families of Equifrequency Transitions}

Perondi \cite{answer} found infinitely many solutions to the generalization of (\ref{main}):
\begin{equation}
   \boxed{ \frac{1}{\beta_1^2}-\frac{1}{\alpha_1^2} = \frac{1}{\beta_2^2}-\frac{1}{\alpha_2^2} = \dots = \frac{1}{\beta_n^2}-\frac{1}{\alpha_n^2}} \ ,
    \label{Perondi}
\end{equation}
for any $n\in \mathbb{Z}_{\ge 2}$. His approach is to start with a set of $k$ primes $S_k = \{\mu_1,\dots,\mu_k\}$, for some $k\in \mathbb{Z}_{\ge 2}$  and then try to find an integer $\Delta$ for which \begin{equation}
    \frac{1}{\beta_1^2}-\frac{1}{\alpha_1^2} = \frac{1}{\beta_2^2}-\frac{1}{\alpha_2^2} = \dots = \frac{1}{\beta_n^2}-\frac{1}{\alpha_n^2} = \frac{4\mu_1\dots \mu_k}{\Delta^2}
    \label{PerondiModified}
\end{equation}
has a solution. By partitioning \cite{halmos1960naive} the set of indices $\{1,\dots,k\}$ into two sets $I,J$ and denoting $\gamma_I = \prod_{i\in I} \mu_i , \gamma_J = \prod_{j\in J} \mu_j$, he found that if $\gamma_I\neq \gamma_J$ and $\Delta_{I,J}$ is divisible by $\gamma_I-\gamma_J$ and $\gamma_I+\gamma_J$, then $(\alpha,\beta) = (\frac{\Delta_{I,J}}{\gamma_I-\gamma_J},\frac{\Delta_{I,J}}{\gamma_I+\gamma_J})$ is a solution to
\begin{equation}
    \frac{1}{\beta^2}-\frac{1}{\alpha^2} = \frac{4\mu_1\dots \mu_k}{\Delta_{I,J}^2} \ ,
    \label{PerondiPartial}
\end{equation}
which again using the identity $(\alpha+\beta)^2-(\alpha-\beta)^2 = 4\alpha\beta$.
Now by enlarging (or shrinking) $S_k$ if necessary and splitting the set of indices differently, he found $n$ distinct pairs $(\gamma_I-\gamma_J,\gamma_I+\gamma_J)$. Then he chose a positive integer $\Delta$ which is divisible by $n$ $\Delta_{I,J}$'s, and found $n$ pairs $(\alpha_{I,J},\beta_{I,J}) = (\frac{\Delta}{\gamma_I-\gamma_J},\frac{\Delta}{\gamma_I+\gamma_J})$ solution of (\ref{PerondiModified}). 

Similar to what we did above, we can find all solutions to the generalized equation (\ref{Perondi}) of Perondi by simply solving the first equation equal the $i^{th}$ equation, simultaneously for all $i$, using the method we found on the previous section. First picking $s\in \mathbb{Z}^{+}$ and $n$ distinct triple $(x_i,y_i,z_i)$ satisfying:
\begin{equation}\begin{cases}
x_1^2 - y_1^2 = s z_1^2 \\
x_2^2 - y_2^2 = s z_2^2 \\
\dots \\
x_n^2 - y_n^2 = s z_n^2 \ .
\end{cases}
\end{equation}
Then we would want to find $t_i$ such that 
\begin{equation} 
\begin{cases}
x_1y_1t_1z_2 = x_2y_2t_2z_1 \\
x_1y_1t_1z_3 = x_3y_3t_3z_1 \\
\dots \\
x_1y_1t_1z_n = x_ny_nt_nz_1 \ .
\end{cases}
\label{productGen}
\end{equation}
It suffices to have $t_1$ to be divisible by 
\begin{equation}
    \left\{\frac{x_iy_iz_1}{\text{gcd}(x_1y_1z_i,x_iy_iz_1)} \text{     for all } 2\le i\le n \right\} \ .
\end{equation}
and then the remaining $t_i$ are deduced from $(\ref{productGen})$. The final solution to the generalized equation (\ref{Perondi}) is:
\begin{equation}
    (\beta_i,\alpha_i) = (y_it_i,x_it_i) \text{     for all } 1\le i\le n \ .
\end{equation}

\vspace{6pt} 

\bibliography{main}
\bibliographystyle{apsrev4-2}

\end{document}